\documentclass[aps,pre,twocolumn,showpacs,showkeys,superscriptaddress]{revtex4}

\usepackage{isolatin1,graphicx,amsmath}
\usepackage[hypertex]{hyperref}

\newcommand{\D}{\mathrm{d}}
\newcommand{\E}{\mathrm{e}}
\newcommand{\dH}{{d_\mathrm{H}}}

\begin{document}

\title{A minimal stochastic model for influenza evolution}

\author{F. Tria}
\affiliation{Abdus Salam ICTP, Strada Costiera 11, I--34100
Trieste (Italy)}
\author{M. Lässig}
\affiliation{Institut für theoretische
Physik, Universität zu Köln, Zülpicher Straße, D-50937 Köln (Germany)}
\author{L. Peliti}\thanks{Associato INFN,
Sezione di Napoli}\email[Corresponding author. E-mail: ]{peliti@na.infn.it}
\affiliation{Dipartimento di Scienze Fisiche and Unità INFM,
Università ``Federico II'', Complesso Monte S. Angelo,
I--80126 Napoli (Italy)}
\author{S. Franz}
\affiliation{Abdus Salam ICTP, Strada Costiera 11, I--34100
Trieste (Italy)}

\date{May 18, 2005}

\begin{abstract}
We introduce and discuss a minimal individual-based model for
influenza dynamics. The model takes into account the effects
of specific immunization against viral strains, but also
infectivity randomness and the presence of a short-lived
strain transcending immunity recently suggested in the
literature. We show by simulations that the resulting model
exhibits substitution of viral strains along the
years, but that their divergence remains bounded.
We also show that dropping any of these features
results in a drastically different behavior, leading
either to the extinction of the disease, to the proliferation
of the viral strains, or to their divergence.
\end{abstract}



\maketitle

\section{Introduction\label{intro}}
Influenza~\cite{Earn} exhibits two apparently contradictory features:
on the one hand,
any given individual can get infected with the disease over and over again,
since the virus mutates fast enough to escape
acquired immunity; on the other hand, on any
given epidemic season, the viral strain is sufficiently
well-defined, so that an effective vaccine can be identified.
The fact that circulating viral strains are closely related
is also exhibited by the shape of its phylogenetic trees~\cite{Fitch1,Fitch2}.
Influenza A, the most relevant epidemiologically,
can be distinguished in several subtypes,
according to the nature of their capside proteins
hemagglutinin (H) and neuroaminidase (N). The currently
prevailing strains belong to the H3N2 subtype.
The phylogenetic relationships of different strains within a subtype are
usually reconstructed from the hemagglutinin (HA) sequences,
since this protein appears to be highly prone to substitutions.
The resulting tree has a characteristic comb-like shape,
with a well-defined backbone and several short-lived side branches.
This has been contrasted with the trees of other viruses,
like HIV or the measles virus,
which show more ramified patterns~\cite{Grenf04}.

This problem has been recently addressed by
Ferguson et al.~\cite{Ferg}, who identified a short-term
strain-transcending immunity as the essential factor
to avoid the dichotomy between extinction and strain proliferation,
and obtained phylogenetic trees quite similar to
the one observed. However, the model
proposed in ref.~\cite{Ferg} contains a very high
number of parameters, including a nontrivial source of
heterogeneity among individuals in the geographic
pattern of transmission.

We build up in this paper what we believe is a minimal model
explaining this feature of influenza epidemiology within a
subtype. The model takes into account the genetic drift of the
virus and the effects of specific immunization. It also 
features a short-term cross-immunity 
effective against all viral strains
(\textit{short immunity}), introduced in ref.~\cite{Ferg}, and the
competition between viral strains, not previously considered, due
to their different infectivities. 
The possible relevance of such an effect is supported by
a recent study~\cite{Moya} which shows that single-nucleotide
substitutions can lead to large fitness changes in RNA viruses.
The removal of any of these features from the model would impair
its viability.

In section~\ref{models} we provide a brief review of
recent models describing the dynamics of influenza.
In section~\ref{model} we describe the model we propose.
An analysis of its behavior and the simulation results
are expounded in section~\ref{results}. We close by a
discussion of our results and an outlook on further research.

\section{Models of influenza dynamics\label{models}}
Recent models of influenza dynamics combine the classical ideas of
mathematical epidemiology with aspects of evolutionary genetics.
In a class of models, represented, e.g., by
\cite{Bornh1,Bornh2,Andrea,Boni}, one assumes the existence of a
single preferred strain at each season, at a given genetic
distance from the previously preferred one. These models show how
the virus population can drift fast enough to remain close to the
preferred strain: however, they do not address the crucial
question mentioned above, namely the quasi-one dimensional
structure of phylogenetic tree.

In a second class of models~\cite{Girvan,Gog,Lin} the genetic
drift of the virus is assumed to take place in a low-dimensional
space ($d=1,2$) with only one viable mutation resolving the
compromise of maintaining the effectiveness of the HA protein and
escaping the detection by the immune system.  These models can be
studied quite deeply with combined analytical and numerical
methods. They exhibit a regime with travelling waves, describing a
persisting genetic drift of the virus with a bounded diversity.
Although from a pragmatic point of view these models provide a
reasonable representation of the observations, directionality in
evolution is assumed rather than derived. This has different
problems: Simple stochastic variations of the model do not lead
to the desired behavior.  A stochastic process that has on average
only one escape direction at a given time would face extinction
after a finite number of steps. Moreover, a detailed analysis of
HA sequences in ref.~\cite{cluster1,cluster2} identifies a non
trivial structure of clusters that succeed one another in time
with abrupt jumps in protein Hamming distance from one cluster to
another. This gives circumstantial evidence that a
larger than one-dimensional manifold in genomic space is involved
in the process.

The most structured attempt to derive a working model of interaction
between viral strain evolution and epidemiological dynamics is
represented by ref.~\cite{Ferg}. In this model a complex spatial
structure of the host population as well as a detailed parametrization
of the dynamics of infection and recovery is introduced, and the viral
evolution takes place in a high-dimensional genomic space. The
main result of the authors is that in order to avoid the proliferation
of strains it is necessary to assume that infection to a given virus,
in addition to conferring long term specific immunity to close strains, it
also elicits a \textit{short immunity} against all possible variants.
Unfortunately the models contains a high number of parameters
and it is difficult to isolate this mechanism from the different
details of the model.

\section{Building up the model\label{model}}
We use an individual-based model generalizing the bitstring model
introduced in ref.~\cite{Girvan}. We consider a population of $N$
individuals, which can be host to the virus. The antigenic
features of the virus are summarized in a binary string $\sigma$
of length $L$. Each host can be in one of the following states:
\begin{description}
\item[Healthy:] The host can be infected by suitable strains of
the virus, depending on its acquired immunity; \item[Infected:]
The host is infected by a unique viral strain $\sigma$;
\item[Recovered:] The host is not infected and cannot be infected
by any viral strain. This state represents those individuals which
are protected by the \textit{short immunity} against infection by
any influenza strain.
\end{description}
The immunity acquired by any host $i$ in its lifetime is described
by the temporally ordered set $\Sigma_i$ of strains which have
infected it in the past. A viral strain $\sigma$ cannot infect a
host $i$ if the set $\Sigma_i$ contains one or more strains
$\sigma'$ such that the Hamming distance $\dH(\sigma,\sigma')\le
r$, where $r$ is the immunity range. Individuals are removed with
rate $\lambda^{-1}$, independently of their state (where $\lambda$
is the average lifetime), and replaced by healthy individuals with
a virgin immune state ($\Sigma=\emptyset$). Similarly, the
duration of the illness  and that of the recovered state are
exponentially distributed with averages $\tau$ and $\eta$
respectively. The memory set $\Sigma_i$ has a maximal length
$\ell_0$. If an individual has been infected by more than $\ell_0$
strains, only the most recent $\ell_0$ ones are remembered. Since
at any given time ``too old'' variants are completely extinct and
out of the immunity range relative to the active ones, the
dynamics of the model is independent of $\ell_0$ if this parameter
is large enough. We have used $\ell_0=50$ in the simulations, but
independence is already reached for $\ell_0\simeq 20$. The
recovered state represents the \textit{short immunity} introduced
by Ferguson et al.~\cite{Ferg}. The disease is transmitted through
random encounters between infected and healthy individuals,
assuming homogeneous mixing, in the
dynamical process that we now describe.

At each time step (representing a duration $\D t/N$) an individual
$i$ is picked up at random. If the individual is infected, one
first checks for possible mutations of the virus: with probability
$\mu \D t$ the state of one of the bits of its strain $\sigma_i$
is changed. Then one picks up at random $\nu$ different
individuals in the population, where $\nu$ is Poisson distributed
with average $\beta(\sigma_i) \D t/\tau$, where $\beta(\sigma_i)$
is the infectivity of the viral strain $\sigma_i$. If one of these
individuals is healthy, and its immune memory does not elicit
immunity, it becomes infected with strain $\sigma_i$. Then,
with probability $\D t/\tau$, the individual $i$ goes to the
recovered state, and the strain $\sigma_i$ is added to its immune
memory $\Sigma_i$. If the individual $i$ is recovered, then it
moves to the healthy state with probability $\D t/\eta$. The
infectivities $\beta(\sigma)$ are assumed to be independent,
identically distributed random variables for each strain $\sigma$,
with a gamma distribution of average $\beta_0$ and parameter $k$:
\begin{equation}
\label{gamma:eq}
p(\beta;k,\beta_0) \propto \left(\frac{k \beta}{\beta_0}\right)^{k-1}\,
\E^{-k \beta/\beta_0}.
\end{equation}
The gamma distribution has the advantage of being easy to
implement and of being naturally defined only for nonnegative
values of $\beta$.

The values of the parameters are chosen to be close to realistic
estimates. One of the problems we meet is to consider populations
large enough to avoid extinctions due to stochastic fluctuations,
and to reasonably implement the immune memory. We could simulate
in reasonable time systems of size up to $N=500000$. We choose the
year as unit of time and set the average duration $\tau$ of the
illness to 0.02 (i.e., roughly one week). We set the average
duration $\eta$ of the recovered state as 0.75 (i.e, after 6
months one has $\sim 50\%$ probability of being no more immune).

According to ref.~\cite{Drake}, the spontaneous genomic mutation
rate $\mu_\mathrm{g}$ for influenza~A viruses equals roughly one
mutation per genome per replication. Taking into account the
duration of a viral generation (a few hours) and the fact that we
are looking at a small portion of the genome ($L=32$ in the
simulations that follow) we can set $\mu\simeq 1$ per strain per
year as a good order of magnitude.

We implemented two versions of this model, that differ by the
duration of the elementary time step. This difference affects the
details of the dynamics, but not the overall behavior of the
model. In the slow version, the elementary time step $\D t$ was
taken as small as 0.001, i.e., about 8 hours. In the fast version,
we took $\D t=\tau$ (the duration of the illness), i.e., at the
end of the infection process, the infected individual was
systematically moved to the recovered state. The fast version was
used to explore the phase space of the model, and the behavior of
the system in the interesting regime was then analyzed in details
with the slow version.

\section{Results\label{results}}
\begin{figure*}
 \includegraphics[width=8cm]{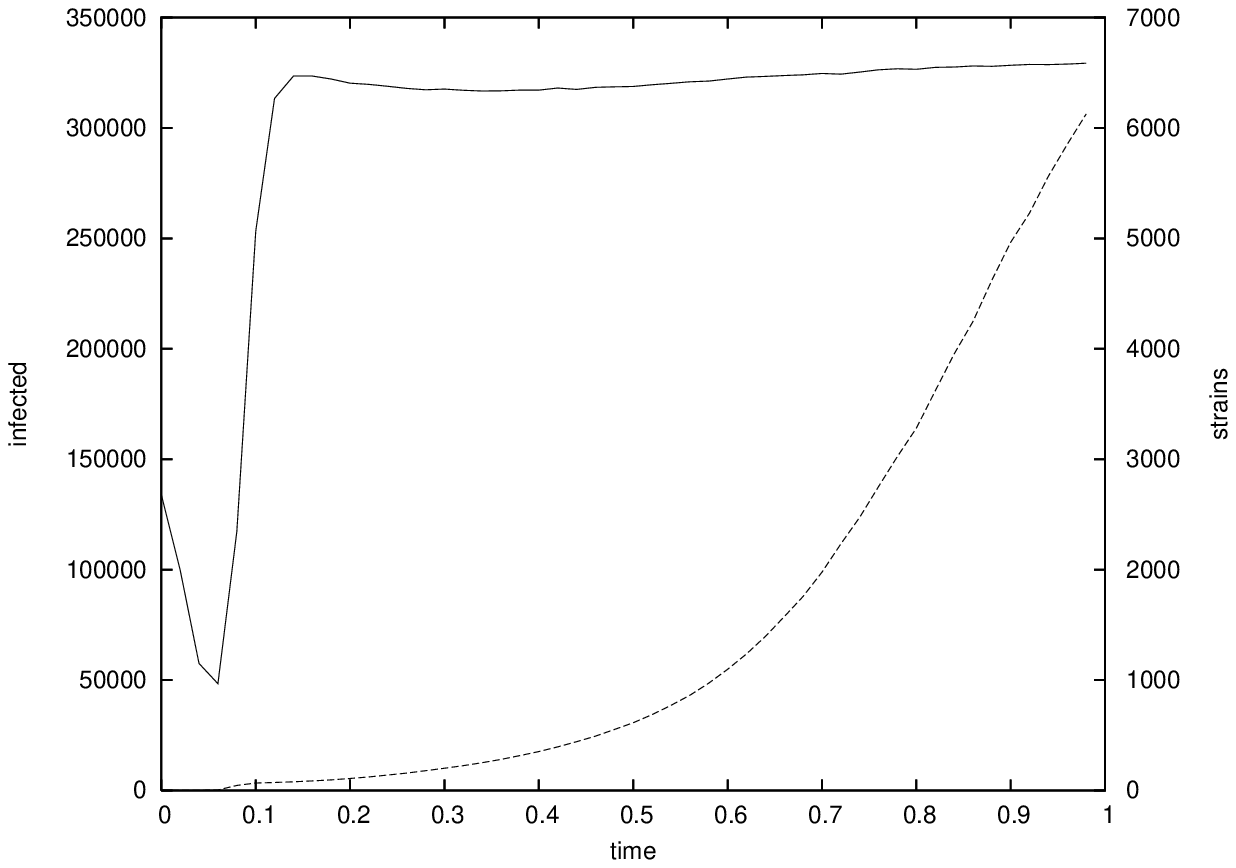}\hspace{0.5cm}
 \includegraphics[width=8cm]{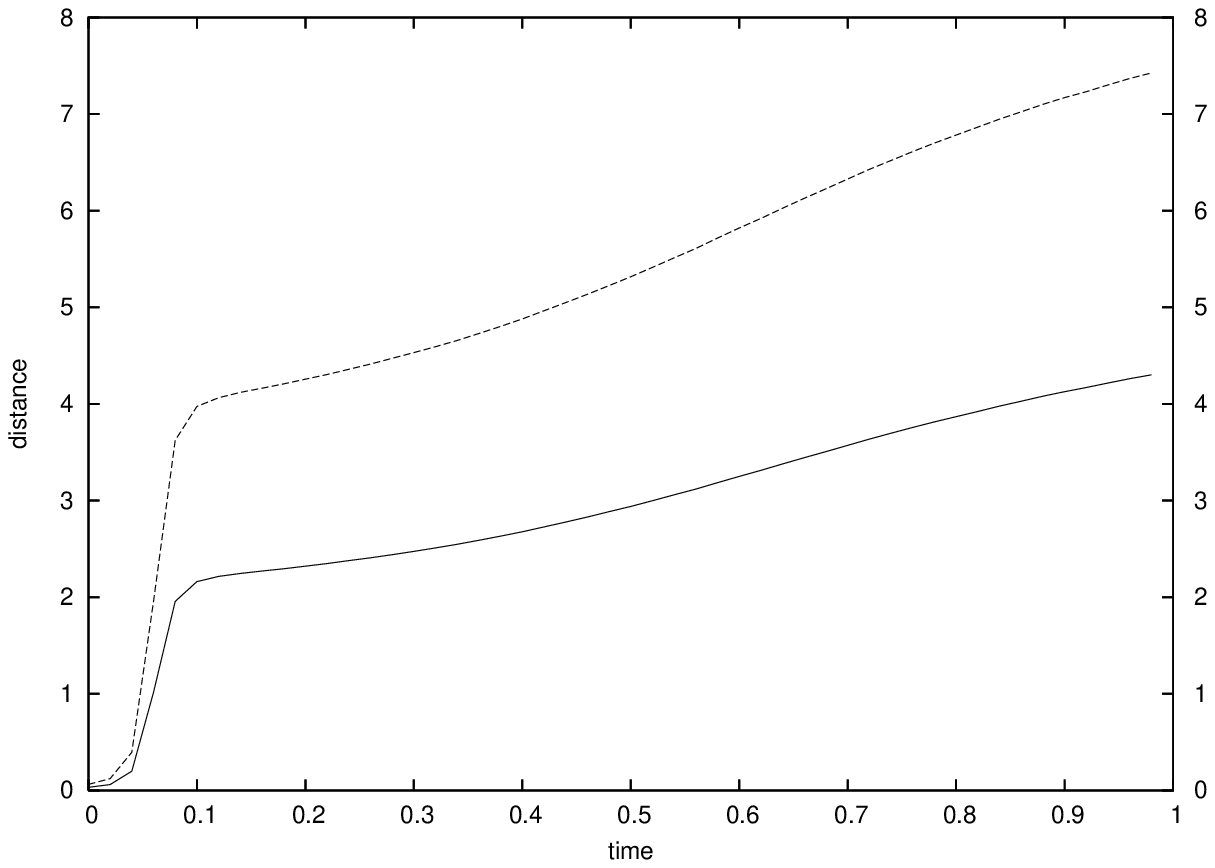}
 \caption{Simulation of the bitstring
 model with 500000 individuals. Parameters: duration
of the illness $\tau=0.02$, lifetime $\lambda=50$,
infectivity $\beta=\beta_0=3$, immunity
range $r=1$. Left:
Continuous line (left axis): number of infected as a function of
time.
Dashed line (right axis): effective number of strains vs.\ time.
Right: Continuous line: Average Hamming distance of the active strains
from the origin. Dashed line: Average Hamming distance between active
strains.
\label{girv1}}
\end{figure*}

We first look at the simple version of the model, in which
the duration of the recovered state is negligible, and the
infectivity $\beta$ is not random. This model coincides with
the bitstring model introduced by Girvan et al.~\cite{Girvan},
and exhibits only two possible behaviors: extinction of the
viral disease, or proliferation of the viral strains.
We show in fig.~\ref{girv1} the results of the simulation
of the model with a population of 500000 individuals.

All results quoted in this explorative
stage were obtained with the fast version of
the dynamics.

An important parameter characterizing the viral population at a
given time is the effective number of circulating strains
$n$,defined as the inverse participation ratio of the numbers
$\nu_\sigma$ of individuals infected by strain $\sigma$, namely
\begin{equation}
\label{straineff}
n=\left(\sum_\sigma \nu_\sigma\right)^2\Big/
\left(\sum_\sigma \nu_\sigma^2\right).
\end{equation}
Since due to mutations one can have many different strains, each
infecting only a small number of individuals, the effective number
of strains can be much smaller than the total one. We also
monitored some quantities which could give us some insight into
the way the viral strains explore sequence space. The system is
initialized with a single viral strain
$\sigma=\sigma_0=(0,0,\ldots,0)$, and mutations set to 1 some of
these zeros. We could thus evaluate, at least initially, the drift
of the strain, by computing the average of the Hamming distance of
the active strains from the initial point:
\begin{equation}
\label{average}
\Delta_\mathrm{H}=\frac{1}{I}\sum_\sigma \dH(\sigma,\sigma_0)\,\nu_\sigma.
\end{equation}
Here $I$ is the incidence of the disease, i.e., the total number of infected:
\begin{equation}
I=\sum_\sigma \nu_\sigma.
\end{equation}
We call ``active'' strains those for which $\nu_\sigma >0$.
On the other hand, the cloud of active viral strains
broadens as it drifts. Its width can be estimated by
evaluating the average mutual Hamming distance between active
strains, weighted by their incidence:
\begin{equation}
\label{mutual}
\delta_\mathrm{H}=\frac{2}{I(I-1)}\sum_{(\sigma,\sigma')}
\dH(\sigma,\sigma')\,\nu_\sigma\nu_{\sigma'},
\end{equation}
where the sum runs over all distinct pairs $(\sigma,\sigma')$
of active strains.

It is interesting that in this regime, as already noticed in
ref.~\cite{Girvan}, the disease gets extinct at large values of
the infectivity. With our data, e.g., for $r=2$, $\mu=1$,
extinction occurs for $\beta_0 < 2$, but also for $\beta_0 > 8$.
This effect is due to the fact that the initial ripple in the
number of infected individuals is followed by a severe bottleneck,
as shown in fig.~\ref{girv1}. The bottleneck becomes more intense
as the infectivity increases, eventually leading to extinction.
The eventual proliferation of viral strains makes it difficult to
analyze this model by generalizing the occupation number
representation framework used, e.g., in~\cite{Gog} to a
multidimensional sequence space~\cite{Tria}.

Since this model cannot sustain the nonproliferating strain regime
characteristic of influenza, we considered if infectivity
randomness alone could be responsible for it. Indeed, recent
studies~\cite{Moya} have shown that single-nucleotide
substitutions lead to a wide distribution of fitness effects in an
RNA virus. It is likely that some of these effects also arise in
the short nucleotide sequence coding for the immunologically
relevant section of NA. We have thus attached to each strain
$\sigma$ a value $\beta_\sigma$ of the infectivity, drawn from a
gamma distribution of average $\beta_0$ and parameter $k$.

\begin{figure*}
\includegraphics[width=8cm]{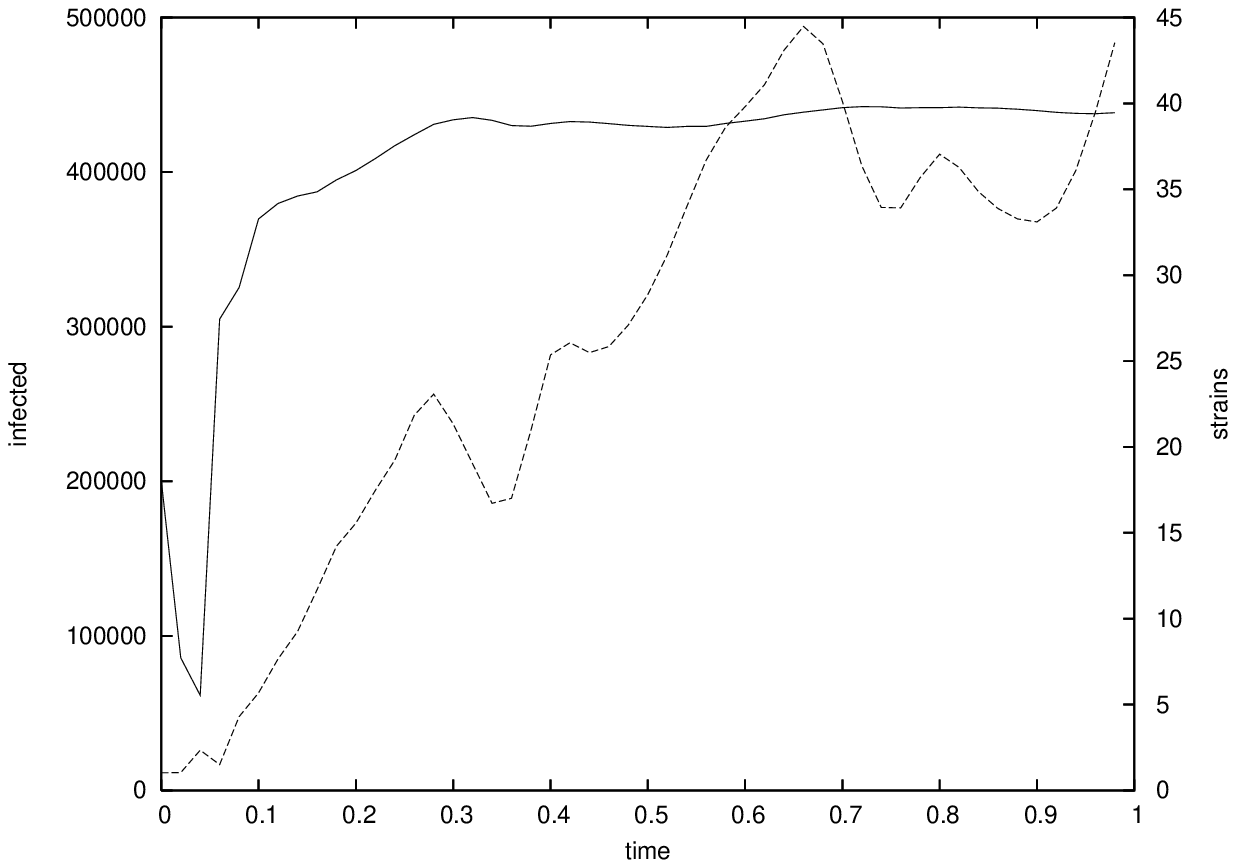}\hspace{0.5cm}
\includegraphics[width=8cm]{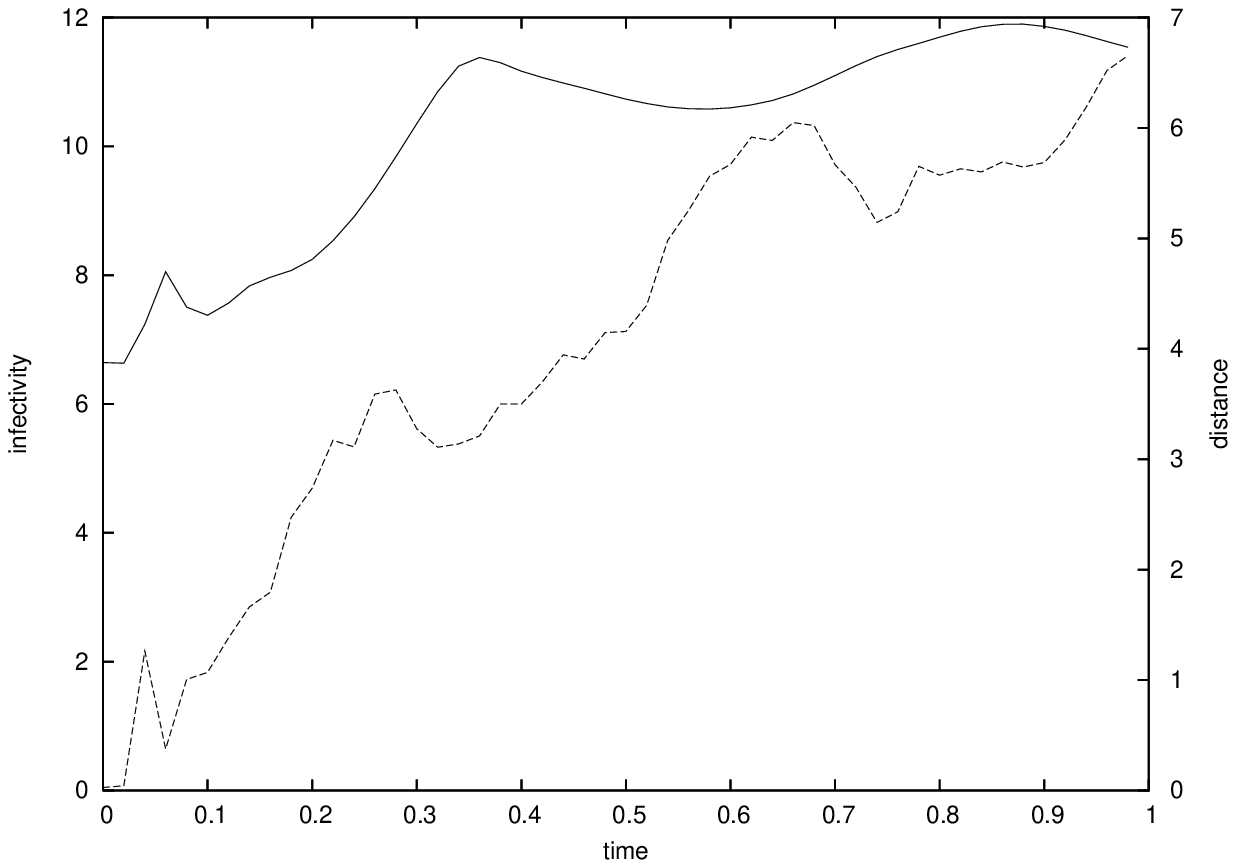}
\caption{Behavior of the system with randomly distributed
infectivity. Expected value of the infectivity $\beta_0=3$, parameter $k=3$,
other parameters as in fig.~\ref{girv1}.
Left: Continuous line (left axis): Number of infected vs.\ time.
Dashed line (right axis): Effective number of strains.
Right: Continuous line (left axis): Average value of the infectivity.
Dashed line (right axis): Average Hamming distance among
active strains. Notice the proliferation and
the divergence of the strains.\label{noferg1}}
\end{figure*}

The behavior of the system in the presence only of the randomly distributed
infectivity does not appear much different from that of the simple bitstring
model. There are only the proliferation and the extinction regimes.
If anything, the phase space allotted to the proliferating regime
is reduced, because the high average values of the infectivity.
The illness did not appear to remain, with our mutation rate
and population size, for ranges $r\ge 3$. The average infectivity
values in the population appear much larger than the average
$\beta_0$ of the distribution, suggesting that
the competition takes place at the tail of the distribution.
This behavior does not depend strongly on the parameter $k$,
although the average values of the infectivity become
smaller as $k$ is increased.

\begin{figure*}
\includegraphics[width=8cm]{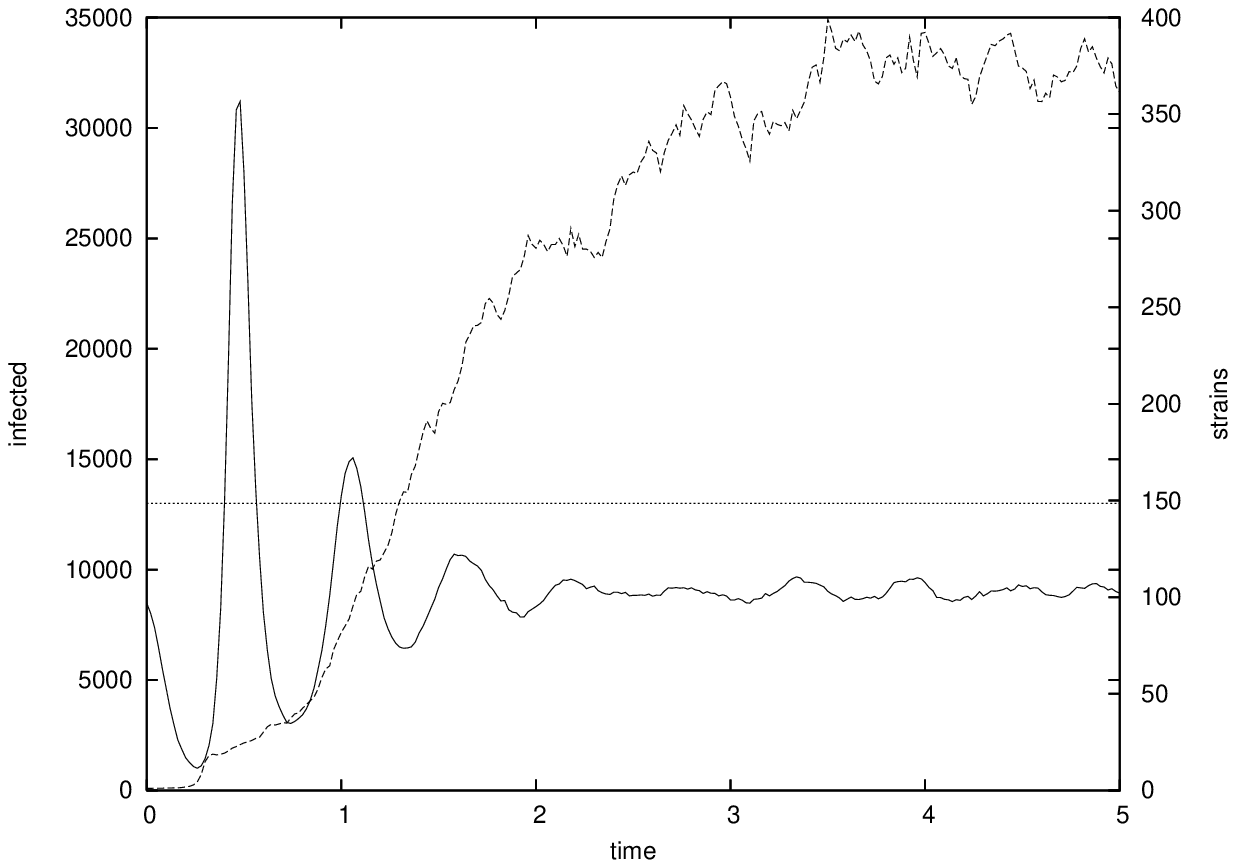}\hspace{0.5cm}
\includegraphics[width=8cm]{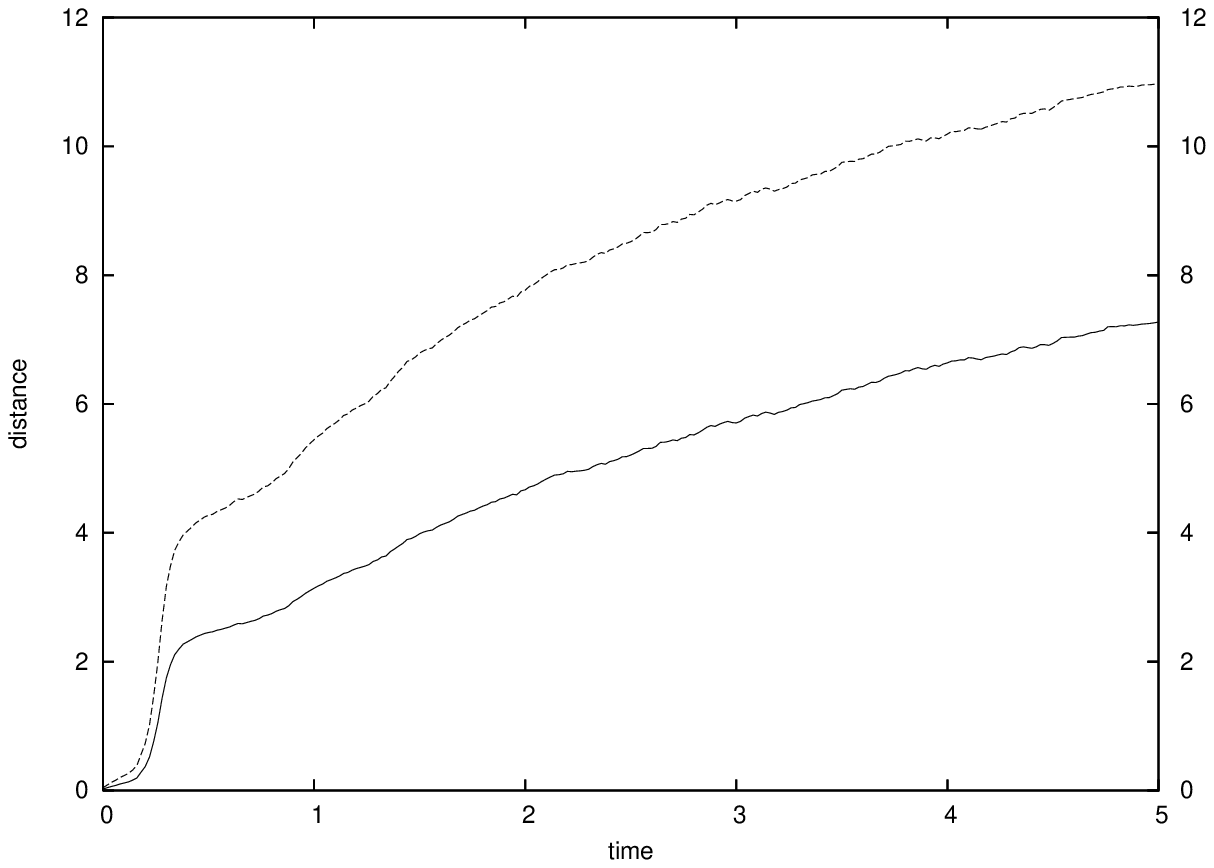}
\caption{Behavior of the system with \textit{short
immunity} but nonrandom infectivity. Infectivity $\beta=\beta_0=3$,
short immunity duration $\eta=0.75$. Other parameters as in fig.~\ref{girv1}.
Left: Continuous line (left axis): Number of infected vs.\ time.
Horizontal line: expected number of infected according
to eq.~(\ref{stationary}).
Dashed line (right axis): Effective number of strains.
Although the number of infected remains close to the equilibrium
value $I^*$, the effective number of strains reaches
a large value, corresponding to proliferation. Right:
Continuous line: Average Hamming distance from the origin of the active
strains. Dashed line: Average Hamming distance between
active strains. One sees that the distance between active strains
keeps increasing, witnessing their divergence. (With $L=32$, the average
distance in a random sample would be $\dH=16$.)\label{norand1}}
\end{figure*}

If the infectivity is nonrandom, but the
short-term general immunity is present, the disease
either dies off or, after a few ripples,
reaches a steady state at an incidence
level (number of infected) of the order of
\begin{equation}
\label{stationary}
I^*=N \frac{\tau}{\tau+\eta}\left(1-\frac{1}{\beta_0}\right).
\end{equation}
In this regime, however, the effects of specific immunization
are apparent only in the small reduction of the
steady-state incidence level, as seen in fig.~\ref{norand1}, left.
On the other hand, one can see in fig.~\ref{norand1} that
the effective number of active strains remains high,
and most of all that they diverge so that the average
Hamming distance soon gets close to the theoretical
value for a random sample.

\begin{figure*}
\includegraphics[width=8cm]{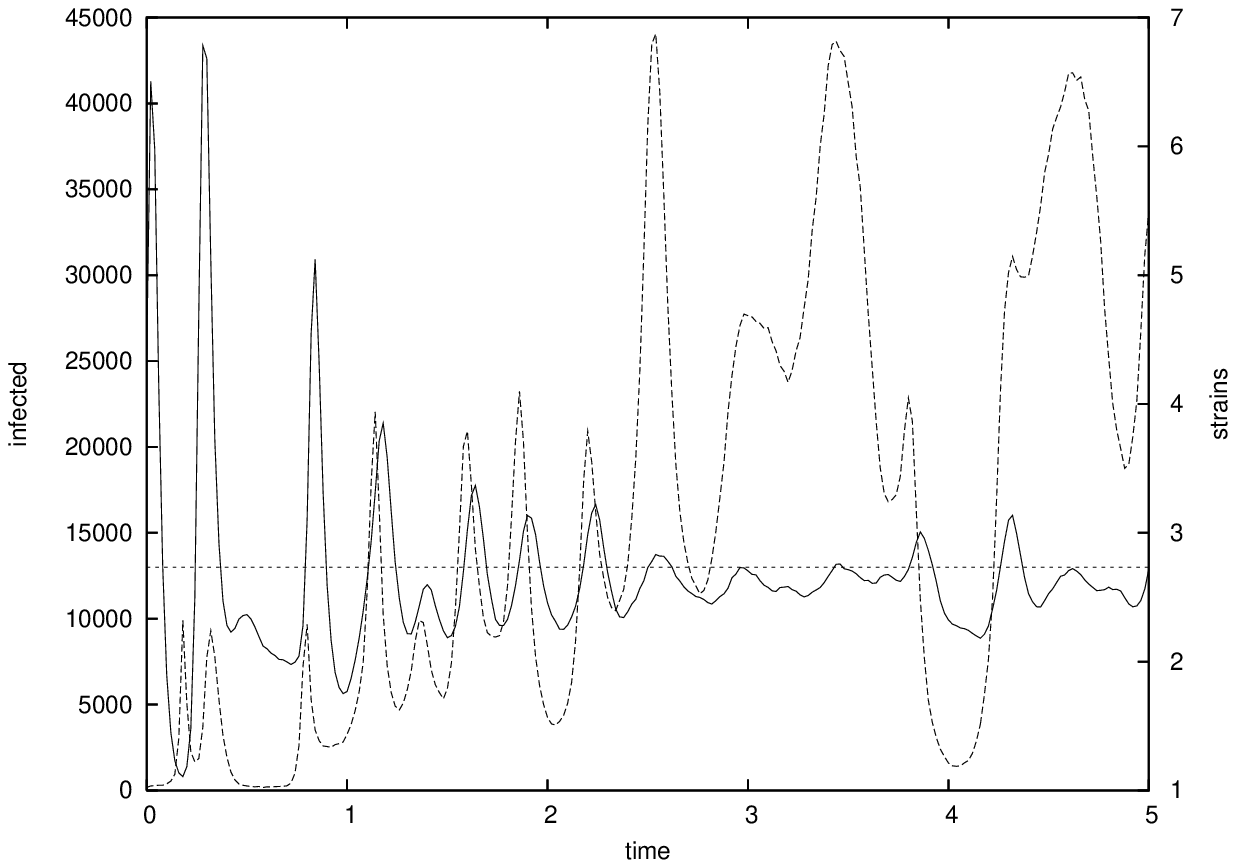}\hspace{0.5cm}
\includegraphics[width=8cm]{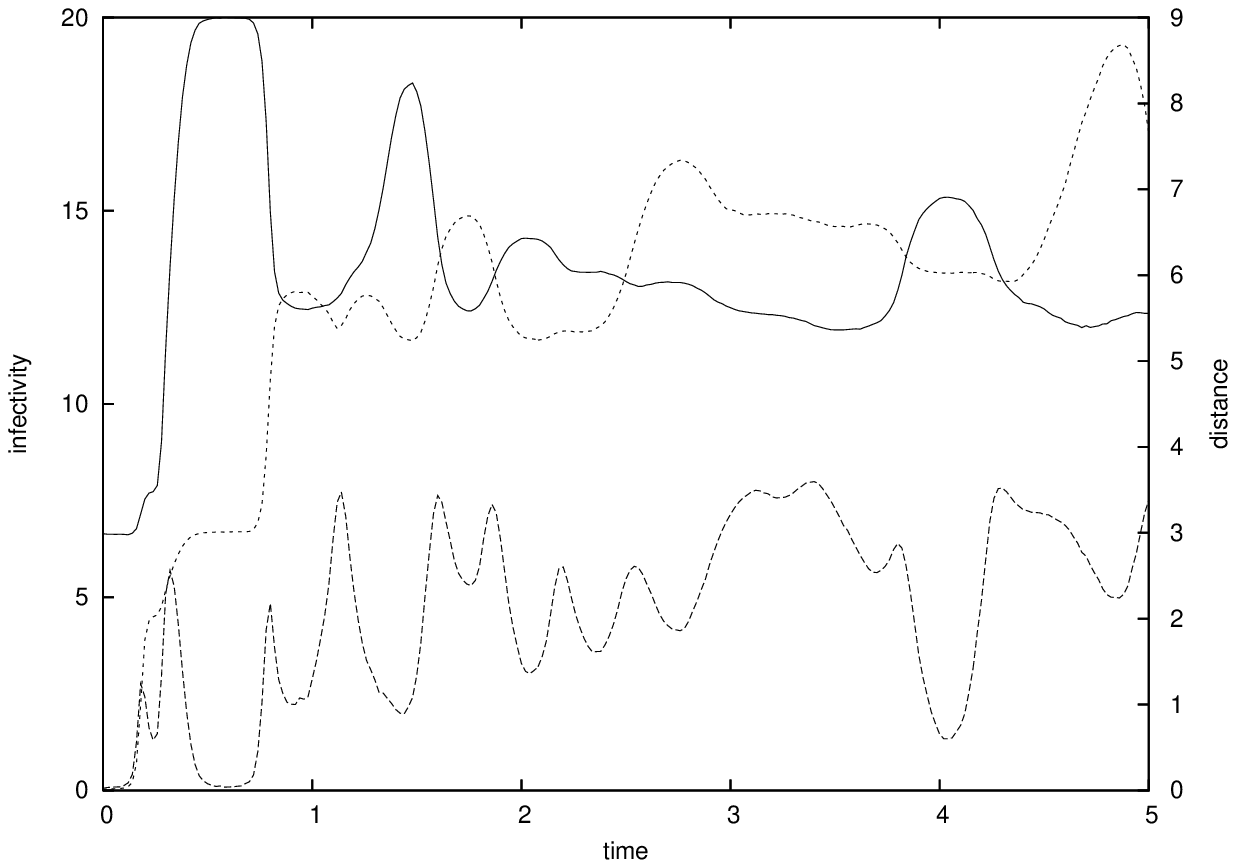}
\caption{Behavior of the system with randomly distributed
infectivity and \textit{short
immunity}. Expected value of the infectivity $\beta_0=3$, parameter $k=3$,
range $r=1$, short immunity duration $\eta=0.75$.
Other parameters as in fig.~\ref{girv1}.
Left: Continuous line (left axis): Number of infected vs.\ time.
Horizontal line: expected number of infected according
to eq.~(\ref{stationary}).
Dashed line (right axis): Effective number of strains.
Right: Continuous line (left axis): Average value of the infectivity.
Dashed line (right axis): Average Hamming distance among
active strains. Dotted line: Average Hamming distance
from the origin for the active strains. Notice that the
number of active strains and their mutual distance
remain limited while exploring sequence space.\label{nomod1}}
\end{figure*}

The picture changes if \textit{both} random infectivity and short immunity
are introduced. We show in fig.~\ref{nomod1} the behavior of
the system with short immunity of duration 0.75 years, and
gamma-distributed infectivity with parameter $k=3$ and an
expected value $\beta_0=3$. Although the incidence of the disease settles
down to a level close to the stationary level dictated
by eq.~(\ref{stationary}), one can see that the effective
number of strains remains of order unity. The competition among
viral strains shows up quite clearly in the oscillating
behavior of $n(t)$ and of $\delta_\mathrm{H}(t)$. One can see
that the actual average values of the infectivity remain
quite high and that, in spite of the ongoing competition, the
incidence of the disease does not show pronounced oscillations.

\begin{figure*}
\includegraphics[width=8cm]{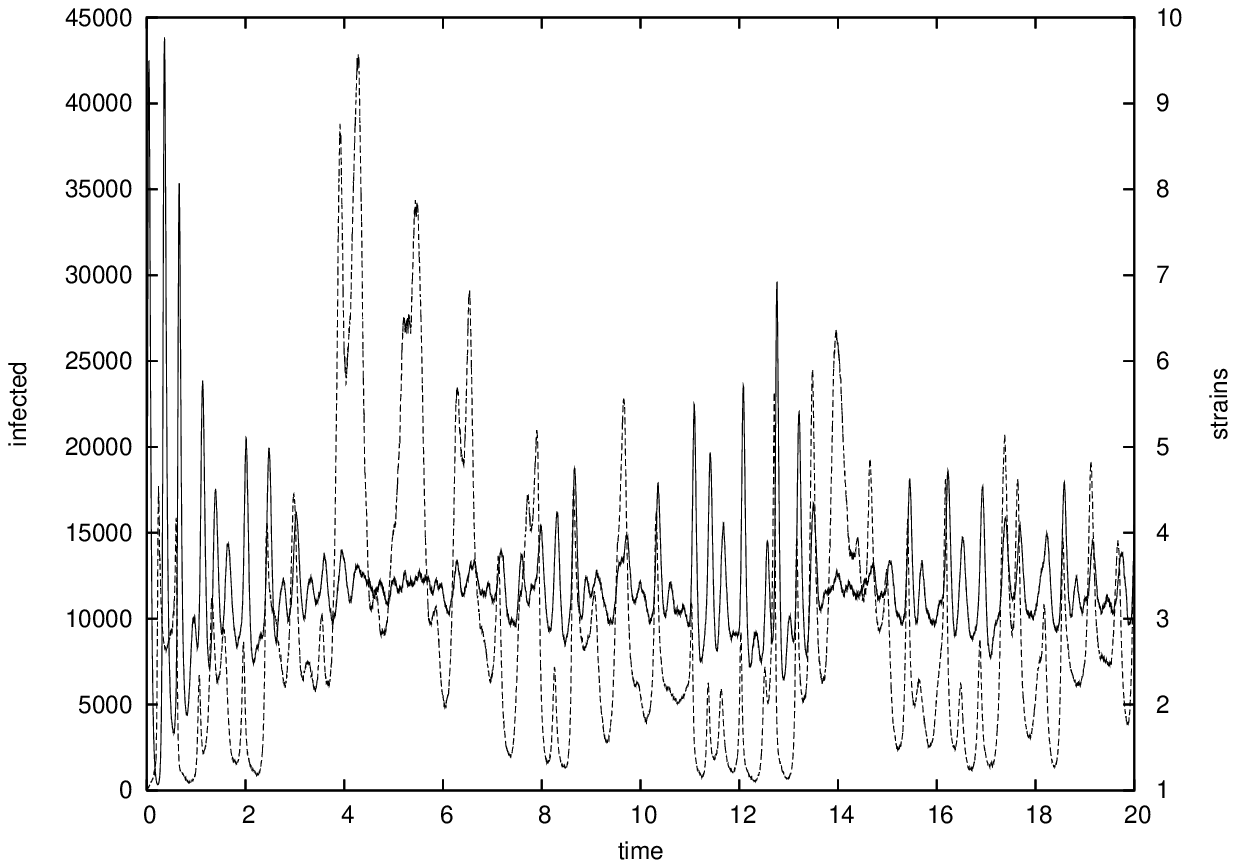}\hspace{0.5cm}
\includegraphics[width=8cm]{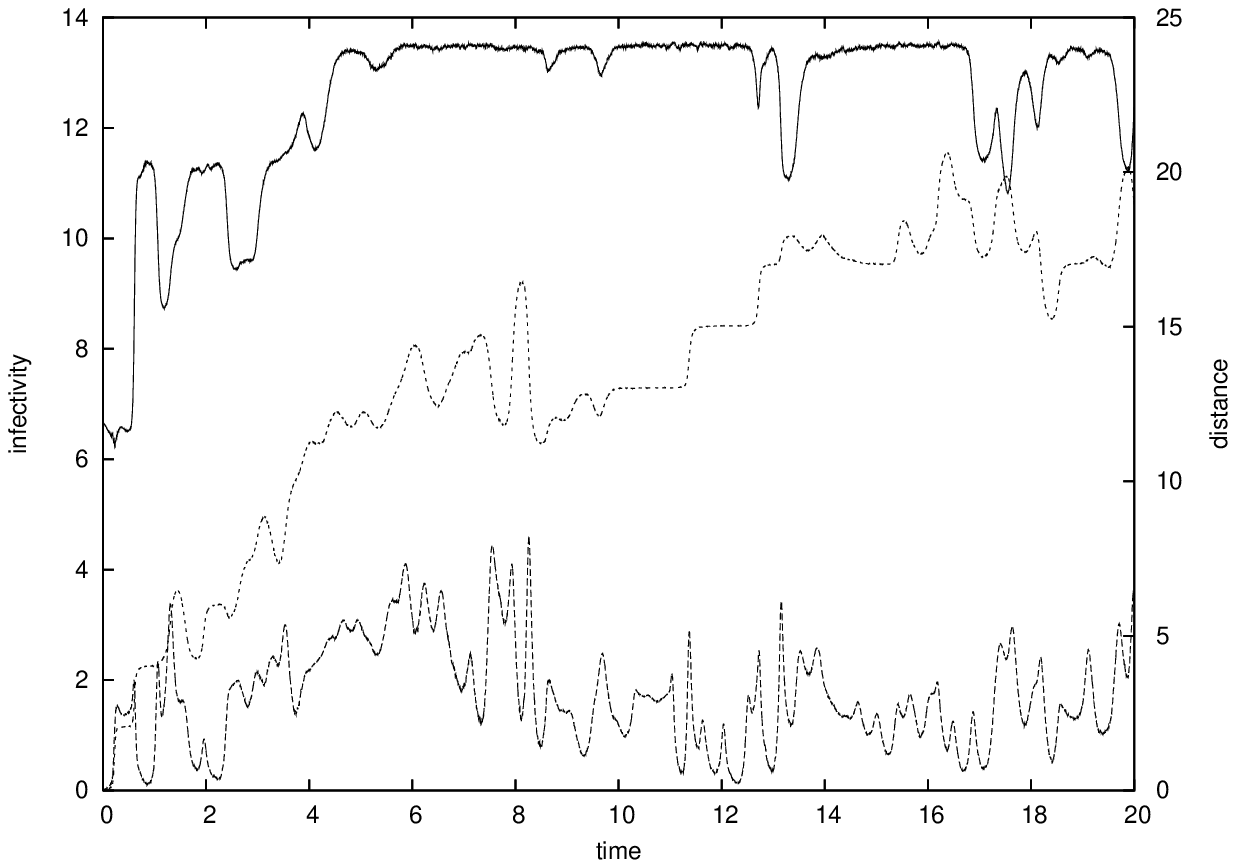}
\caption{Behavior of the slow version of the model
with randomly distributed
infectivity and \textit{short immunity}.
Expected value of the infectivity $\beta_0=3$, parameter $k=3$,
range $r=1$, short immunity duration $\eta=0.75$, genome length $L=128$.
Other parameters as in fig.~\ref{girv1}.
Left: Continuous line (left axis): Number of infected vs.\ time.
Dashed line (right axis): Effective number of strains.
Right: Continuous line (left axis): Average value of the infectivity.
Dashed line (right axis): Average Hamming distance among
active strains. Dotted line: Average Hamming distance
from the origin for the active strains. The incidence ripples are more
evident in this version since the correlations in the
immunity state of the population are more correctly taken
into account.\label{slow1}}
\end{figure*}

We can now consider the slow version of the program, in
order to analyze the behavior of the model in details.
We show in fig.~\ref{slow1} the results of a simulation
in which the time interval $\D t$ is taken equal to
$10^{-3}$~years (corresponding to a few hours), and
the other parameters are as in fig.~\ref{nomod1},
apart from the genome length $L$, which is set equal to
128 in order to reduce the probability of back mutations.
While the overall incidence level remains close to
that previously obtained, one can see that spontaneous
oscillations in the number of infected persist, and are
synchronized with corresponding oscillations in
the effective number of strains and in the width of
the distribution of strains in sequence space.
In this systems, something analogous to the working
regime of influenza appears to have been reached.
One may also notice that the divergence between active strains,
measured by the average Hamming distance, remains limited even if
the average distance from the origin increases with time.
The active strains show therefore ongoing change, but with a
limited amount of divergence.

\begin{figure*}
\includegraphics[width=8cm]{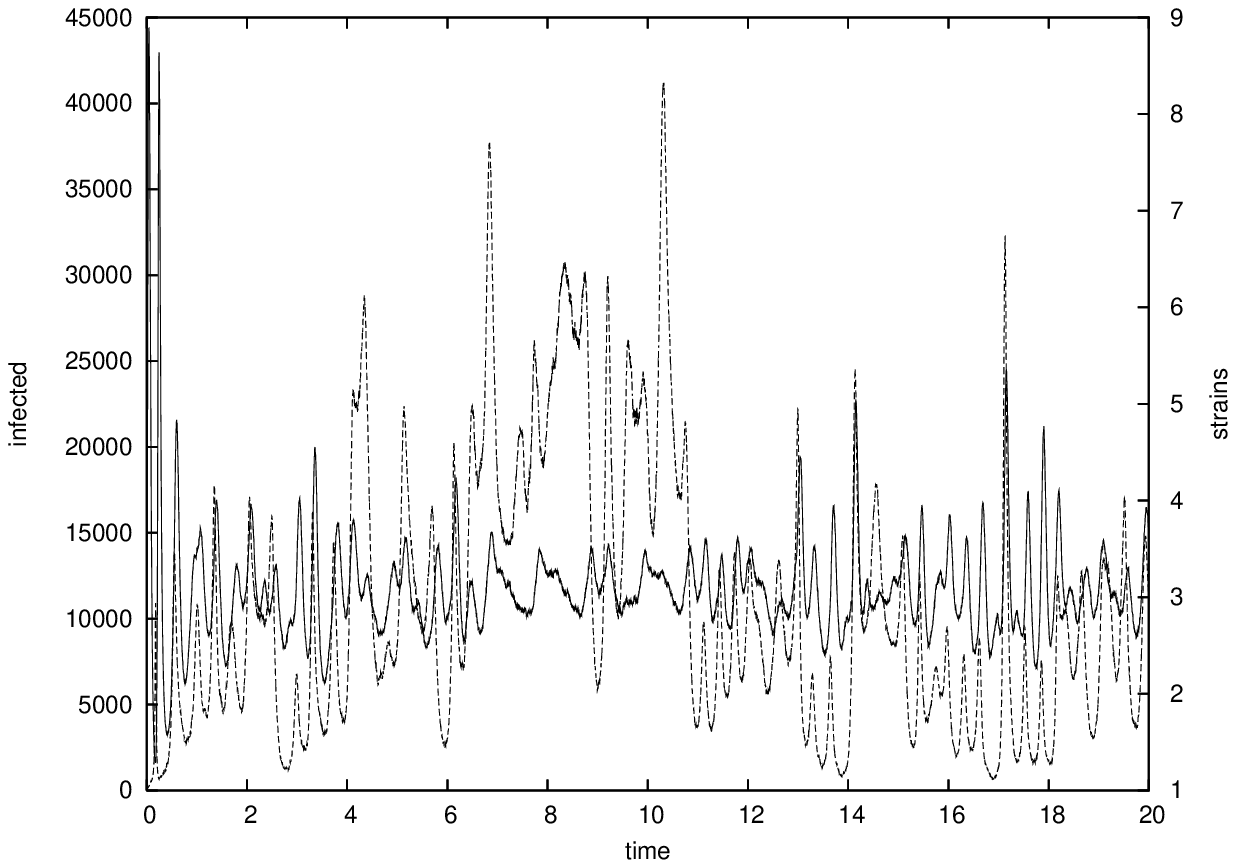}\hspace{0.5cm}
\includegraphics[width=8cm]{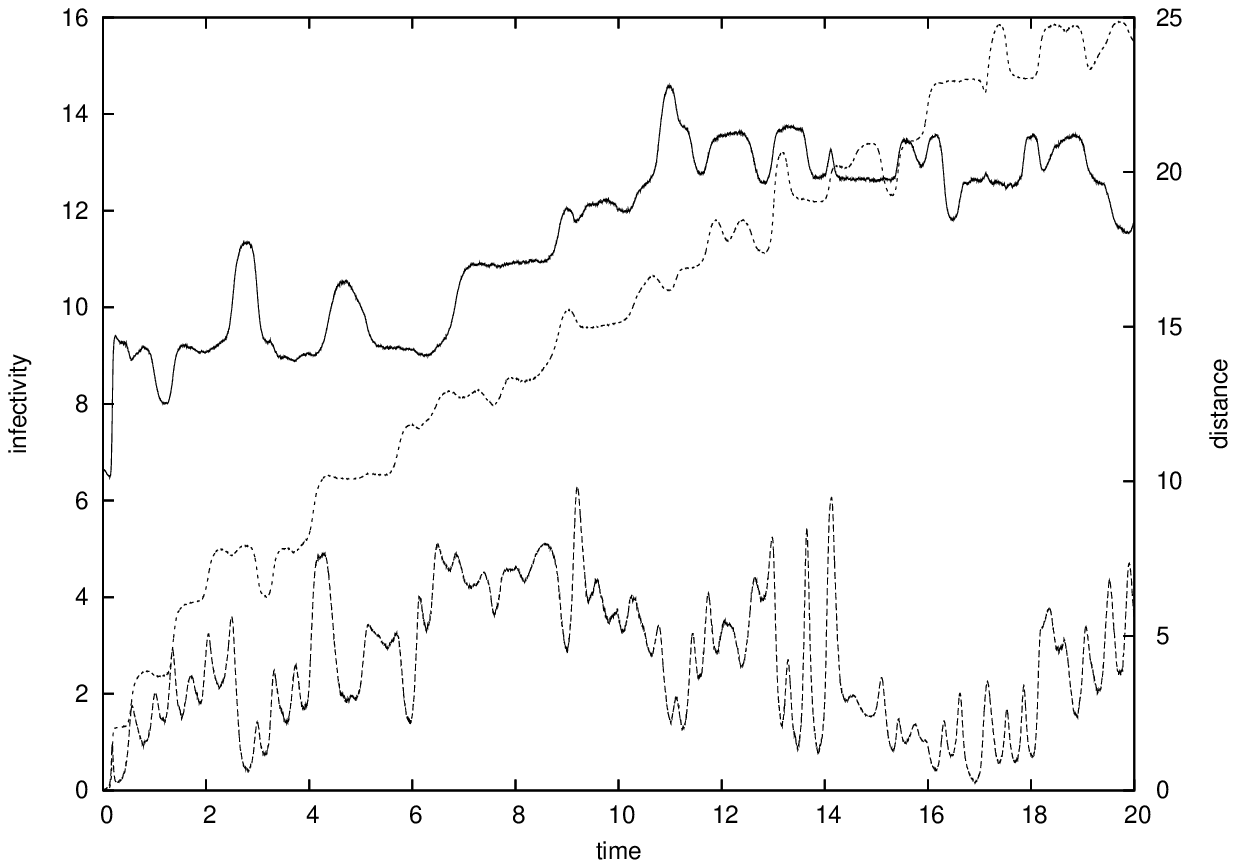}
\caption{Behavior of the slow version of the model
with randomly distributed
infectivity and \textit{short immunity}, in the presence of
a modulation of the infectivity according to eq.~(\ref{modulation}).
Modulation $m_0=0.2$.
Other parameters as in fig.~\ref{slow1}.
Left: Continuous line (left axis): Number of infected vs.\ time.
Dashed line (right axis): Effective number of strains.
Right: Continuous line (left axis): Average value of the infectivity.
Dashed line (right axis): Average Hamming distance among
active strains. Dotted line: Average Hamming distance
from the origin for the active strains. The incidence ripples are more
evident in this version since the correlations in the
immunity state of the population are more correctly taken
into account.\label{slowmod}}
\end{figure*}

These results still hold if one introduces a small modulation
with a period of one year in the infectivity~\footnote{The
possibility that a small modulation in the
infectivity could lead to the observed strong
seasonality of the influenza epidemics is discussed
in~\cite{resonance}.}, by letting,
for each viral strain $\sigma$,
\begin{equation}
\label{modulation}
\beta_\sigma(t)=\beta_\sigma \left(1+m_0 \sin(t/2\pi)\right).
\end{equation}
However, one can see that the characteristic time of the incidence
ripples in our model correspond to a few months, as if it were
essentially determined by the duration of the short immunity and
of the illness, instead of by the viral strain turnover. Therefore
a small modulation does little more than modulate the amplitude of
the ripples, as shown in fig.~\ref{slowmod}.

\section{Discussion}
In this paper we addressed the problem of understanding the
dynamical mechanisms underlying antigenic drift  of the influenza
A virus. In particular we revisited the problem outlined in
\cite{Ferg} of the existence of a constantly evolving well-defined
strain giving rise to comb-like shape phylogenetic trees, i.e., to
a constantly evolving viral population with comparatively narrow
distribution in genetic space. We considered a minimal, individual
based model that couples epidemic dynamics and viral evolution.
Our main finding is that the absence of strain proliferation
relates equally importantly to the large spectrum short-time
cross-immunity emphasized in \cite{Ferg} and to heterogeneity in
the effective viral infectivity. In our model we directly suppose
that different strains have different values of the infectivity.
Our results show that even a comparatively narrow distribution of
infectivity, in combination with the increased competition due to
the presence of the short immunity, is sufficient to lead to a
``drifting quasispecies'' behavior. We have also seen that, on the
other hand, such a behavior has a very narrow range of stability,
if any, if either the short immunity or the random infectivity is
lacking, at least in a model of a comparatively large population
without spatial structure, like the one we have considered. We
expect, on the basis of Kimura's theory of selection in finite
populations~\cite{Kimura}, that, as the population becomes larger
and larger, the spread in viral infectivity needed to stabilize
influenza behavior becomes narrower and narrower.

Perhaps obscured by the emphasis on short time immunity, a
different mechanism was present in~\cite{Ferg}. In that case the
heterogeneity was provided by a non trivial ``geographical
distribution'' of individuals. These were assumed to be randomly
distributed on a two dimensional space. As an effect, the competition
between different viral strains is enhanced, since some geographical
locations acquire a leading role in spreading the disease,
and the first strains to establish themselves in these
locations acquire a standing advantage. However, the possibility
that different strains are also characterized by different
infectivities is quite natural and yields a realistic evolutionary
dynamics. It appears that some heterogeneity and a mechanism for
an enhanced competition among strains is all that is needed to 
reproduce the observed evolutionary pattern. Most probably, varying
infectivities for different strains and a heterogeneous pattern of
contacts among individuals in different communities both play a
role in influenza spreading and viral evolution. We think that
this point deserves further research. Understanding which contact
and social patterns favor an increased effective infectivity could
lead to more effective policies to keep under control influenza
epidemics.

Finally, phylogenetic trees similar to the ones of the influenza
virus have been observed in in-host HIV evolution~\cite{Grenf04}.
The present work might be a stimulus to identify the factors that
play the role corresponding to short immunity and heterogeneous
infectivities in that case.

\begin{acknowledgments}
We are grateful to W. Fitch for discussions and hints. LP thanks the
Abdus Salam ICTP for hospitality. ML and LP are grateful to
the Centro Ennio De Giorgi in Pisa for hosting some of their
discussions.
LP thanks Ester Lázaro for a critical reading of the manuscript.
This work was supported in part by the
European Community's Human Potential programme under contract
``HPRN-CT-2002-00319 STIPCO''. 
\end{acknowledgments}

\bibliography{influenza}

\end{document}